\documentclass[]{amsart}

\newtheorem{lemma}{Lemma}
\newtheorem{theorem}{Theorem}
\newtheorem{corollary}{Corollary}
\newtheorem{hypothesis}{Hypothesis}
\newtheorem{remark}{Remark}
\newtheorem{definition}{Definition}

\newcommand{\asy}{{\it O}}
\def\norma#1{\|#1\|}

\begin{document}

\title []{Stability of spectral eigenspaces in nonlinear Schr\"odinger equations}

\author {Dario Bambusi}
\address {Dipartimento di Matematica\\ Universit\'a degli studi 
di Milano\\ Via Saldini 50, Milano 20133, Italy}

\email {bambusi@mat.unimi.it}

\author {Andrea Sacchetti}

\address {Dipartimento di Matematica Pura ed Applicata\\ Universit\'a 
degli studi di Modena e Reggio Emilia\\ Via Campi 213/B, Modena 41100, Italy}

\email {Sacchetti@unimore.it}

\date {\today}

\keywords {Nonlinear Schr\"odinger equations; almost invariant manifolds.}

\thanks {This work is partially supported by the INdAM project {\it Mathematical modeling and numerical analysis of quantum systems with applications to nanosciences}.}

\begin {abstract}
We consider the time-dependent non linear Schr\"odinger equations with a double well potential in dimensions $d =1$ and $d=2$. \ We prove, in the semiclassical limit, that the finite dimensional eigenspace associated to the lowest two eigenvalues of the linear operator is
almost invariant for any time.
\end{abstract}

\subjclass {Primary {35Bxx}; Secondary {35Q40, 35K55}}

\maketitle

\section {Introduction}
Here we consider the time-dependent nonlinear Schr\"odinger equation (hereafter NLS)
\begin{eqnarray}
\left \{
\begin {array}{l}
i \hbar \dot \psi^t = H_0 \psi^t + \epsilon W \psi^t , \ \ \epsilon \in {\mathbb R}
,\ \dot \psi^t = \frac {\partial \psi^t }{\partial t}, \\ \left. \psi^t (x)\right |_{t=0} =
\psi^0 (x) \in L^2 ({\mathbb R}^d ) , \ \| \psi^0 \|_{L^2} =1 , 
\end {array}
\right. \label {eq1}
\end{eqnarray}
where
\begin{eqnarray} 
H_0 = - \frac {\hbar^2}{2m} \Delta + V , \ \ \ \Delta =
\sum_{j=1}^d \frac {\partial^2}{\partial x_j^2}, \ d= 1 , 2 ,\ \ 2m=1
\label {eq2} 
\end{eqnarray}
is the linear Hamiltonian operator with symmetric double-well potential $V(x)$,
and where
\begin{eqnarray}
W = W (|\psi |^2) = |\psi |^{2\sigma}, \ \ \sigma >0  \label {eq3}
\end{eqnarray}
is a nonlinear perturbation.

We recall that a double well potential is a positive potential symmetric with respect to the reflection through a hyperplane, and having two nondegenerate distinct absolute minima. \ When the nonlinear term is absent, the linear Hamiltonian $H_0$ has the two eigenfunctions corresponding to the two lowest eigenvalues which have either even or odd-parity. \ Given an initial datum which is a linear combination of such lowest eigenfunctions, then the corresponding solution performs a \emph {beating motion}, namely the probability density oscillates periodically from a state almost concentrated on one minimum to a state almost concentrated on the other one. \ The \emph {beating period} usually plays the role of unit of time. \ When the nonlinear term is restored, a symmetry breaking phenomenon occurs: that is, if the strength $\epsilon$ of the nonlinear term is larger than a threshold value then new asymmetric stationary states appears \cite {AFGST}, \cite {GM}, \cite {Zh}. \ Furthermore, for higher strength $\epsilon$ of the nonlinear term, the beating motion is generically forbidden \cite {GMS}, \cite {RSFS}, \cite {VA}. \ These results can be heuristically obtained by reducing the NLS equation to a 2--dimensional dynamical system, namely the restriction of \eqref{eq1} to the space generated by the lowest two eigenvectors. \ Mathematically the difficulty consists in proving that the dynamics of the complete NLS is close to the dynamics of the 2--dimensional reduced system. \ In \cite {Sa2}, making use of semiclassical estimates and refined existence results for NLS this stability result has been obtained for \emph {times of the order of the beating period} in dimensions $d=1,2$ and any $\sigma \in {\mathbb R}^+$.

In the present paper we concentrate on the case of local nonlinearity
\eqref{eq3} with $\sigma$ a positive integer, in this framework we
extend the previous result by \cite {Sa2} proving that, in the
semiclassical limit, the 2-dimensional eigenspace is almost invariant
\emph {for any time} (Theorem \ref {Thm1}). However, due to the
possible presence of positive Lyapunof exponents, our result does not
allow to show that the 2-dimensional system describes the dynamics
over time scales larger than $\epsilon^{-1}$.

Our result is obtained here making use of variational methods and by introducing the scale of Hilbert spaces ${\mathcal X}_s = {\mathcal D} (H_0^{s/2} )$, $s \ge 0$, constituted by the domains of the powers of $H_0$, endowed by the graph norm; in fact, in dimension $d=1$ and $d=2$ we make use of the energy space ${\mathcal X}_1$. 

We close this section by introducing some notations:

\begin {itemize}

\item [-] The notation $y=\asy (e^{-\Gamma /\hbar} )$ means that there exist $\hbar^\star >0$ and a positive constant $C>0$, independent of $\hbar$, such that
\begin{eqnarray*} 
|y| \le C e^{-\Gamma /\hbar}, \ \ \forall \hbar \in (0,\hbar^\star ) 
\end{eqnarray*}

\item [-] The notation $y=\tilde \asy (e^{-\Gamma /\hbar} )$ means that for any $\Gamma '$, $0<\Gamma ' <\Gamma$, then $y=\asy (e^{-\Gamma' /\hbar } )$; that is, there exist $\hbar^\star >0$ and a positive constant $C = C_{\Gamma'} >0$, independent of $\hbar$, such that
\begin{eqnarray*}
|y| \le C e^{-\Gamma' /\hbar} , \ \ \forall \hbar \in (0,\hbar^\star ).
\end{eqnarray*}

\item [-] As usual, $\| \cdot \|_{L^p}$ usually denotes the norm of the space $L^p ({\mathbb R}^d )$; $\| \cdot \|_s$ denotes the norm of the Hilbert spaces ${\mathcal X}_s$, in such a notation ${\mathcal X}_0 =L^2 ({\mathbb R}^d )$ and $\| \cdot \|_0 = \| \cdot \|_{L^2}$.

\item [-] As usual, ${\mathbb R}$ denotes the set of real numbers, ${\mathbb R}^+$ denotes the set of positive real numbers, ${\mathbb N}$ denotes the set of positive integer numbers, $C$ denotes any positive constant independent of $\hbar$ and $t$, and $\langle x \rangle = \sqrt {1+ |x|^2}$, $|x|^2 = \sum_{j=1}^d x_j^2$, where $x =(x_1 , \ldots , x_d) \in {\mathbb R}^d $.

\end {itemize}

\section {The model: preliminary assumptions and main result}

Hereafter we assume the dimension $d=1$ or $d=2$.

\subsection {Double well potentials} 

\begin {hypothesis} 

The double-well potential $V (x)\in C^\infty ({\mathbb R}^d )$ is a symmetric real valued function such that:

\begin {itemize}

\item [i.] $V (x)$ admits two minima at $x=x_\pm $, $x_- \not= x_+$, such that 
\begin{eqnarray*} 
V (x) > V_{min} = V(x_\pm ) , \ \ \forall x \in {\mathbb R}^d , \ x\not= x_\pm ; 
\end{eqnarray*}
for the sake of definitess let us assume
\begin{eqnarray*}
V_{min} =1.
\end{eqnarray*}

\item [ii.] There exists $m>0 $ such that, for large $|x|$ one has 
\begin{eqnarray*}
C \langle x \rangle^m \le V(x) \le C^{-1} \langle x \rangle^m
\end{eqnarray*}
for some constant $0<C<1$;

\item [iii.] For any multi-index $\alpha \in {\mathbb N}^d$ there exists a positive constant $C_\alpha$ such that
\begin{eqnarray*} 
\left | \partial_{x_1}^{\alpha_1} \partial_{x_2}^{\alpha_2} \cdots \partial_{x_d}^{\alpha_d} V(x) \right | \le C_\alpha \langle x \rangle^{m-|\alpha |} 
\end{eqnarray*}

\end {itemize}

\end {hypothesis}

\begin {remark} For the sake of definiteness, we assume that the symmetric potential is such that
\begin{eqnarray*} 
V(-x_1,x_2, \ldots , x_d )= V(x_1,x_2, \ldots , x_d) \ .
\end{eqnarray*}
Furthermore, for the sake of simplicity we assume also that 
\begin{eqnarray*}
\nabla V (x_\pm ) =0 \ \ \mbox { and } \ \ \mbox { Hess } V (x_\pm  ) >0 ; 
\end{eqnarray*}
the case of degenerate minima, that is $ \mbox {det} [\mbox {Hess} V(x_\pm )]=0$, could be treated in a similar way; however, we don't develop here the details.
\end {remark} 

\begin {remark}
As an example one can consider the following double-well potential
\begin{eqnarray*}
V(x)=1+\sum_{j=1}^d \omega_j^2 x_j^2 + V_1 (x) 
\end{eqnarray*}
where $V_1 \in L^\infty ({\mathbb R}^d ) $ is a symmetric function
\begin{eqnarray*}
V_1 (-x_1 , x_2 , \ldots , x_d) = V_1(x_1 , x_2 , \ldots , x_d).
\end{eqnarray*}
\end {remark}

\subsection {Well-posedness of the Cauchy problem.} 

Under Hyp. 1 the Cauchy problem \eqref {eq1}--\eqref {eq3} is locally well posed. 

In fact, in the quadratic (i.e. $m=2$) and sub-quadratic (i.e. $m <2$) cases if $\psi^0 \in H^1$ is such that 
\begin{eqnarray*}
\int_{{\mathbb R}^d} V(x) | \psi^0 (x) |^2 \mbox {\rm d}^d x < \infty
\end{eqnarray*}
then (see \cite {Z} and \S 9.2 in \cite {C}) the Cauchy problem admits a unique bounded local solution $\psi^t \in H^1$ such that 
\begin{eqnarray*}
\int_{{\mathbb R}^d} V(x) | \psi^t (x) |^2 \mbox {\rm d}^d x < \infty
\end{eqnarray*}
for any $t\in [0,\delta )$, for some $\delta>0$. 

Moreover, $\psi^t$ satisfies the following conservation norm
\begin{eqnarray}
{\mathcal N} (\psi^t ) = {\mathcal N} (\psi^0 ) \ ,\label {eq4}
\end{eqnarray}
where 
\begin{eqnarray*}
{\mathcal N } (\psi )= \| \psi \|_{L^2} = \left [ \int_{{\mathbb R}^d} |\psi^0 (x) |^2 \mbox {\rm d}^d x \right ]^{\frac 12} 
\end{eqnarray*}
and 
\begin{eqnarray*} 
{\mathcal E}_\epsilon (\psi^t ) = {\mathcal E}_\epsilon (\psi^0 )
\end{eqnarray*}
where the \emph {energy} functional is defined as
\begin{eqnarray}
{\mathcal E}_\epsilon (\psi ) = {\mathcal E}_0 (\psi ) + \epsilon P (\psi ) \ , \label {eq5}
\end{eqnarray}
where 
\begin{eqnarray*}
{\mathcal E}_0 (\psi ) = \langle \psi , H_0 \psi \rangle_{L^2} \ \mbox { and } \ P(\psi ) =\frac {1}{1+\sigma} \| \psi^{\sigma +1} \|_{L^2}^2. 
\end{eqnarray*}

For super-quadratic potentials (i.e. $m>2$) if $\psi^0 \in {\mathcal X}_s$ then (see Theorem 1.5 in \cite {YZ04}) the Cauchy problem admits an unique solution $\psi^t $ for any $t\in [0,\delta )$, where $\delta$ continously depends on $\| \psi^0 \|_s$. \ Furthermore, the solution $\psi^t$ continuously depends on $t$ and on the initial condition $\psi^0$. \ The conservation of the norm \eqref {eq4} and of the energy \eqref {eq5} will follow from standard density arguments \cite {C} and from the Gagliardo-Nirenberg inequality, which implies, in dimension $d=1$ and $d=2$, the continuity, with respect the norm $H^1$, of the energy functional:
\begin{eqnarray*}
{\mathcal E}_\epsilon (\psi _m ) \to {\mathcal E}_\epsilon (\psi ) , \ \ \mbox { as } \ \psi_m  \stackrel{H^1}{\rightarrow} \psi .
\end{eqnarray*}

Notice that when $\epsilon >0$ then the conservation of the energy implies that $\| \nabla \psi \| \le C$ for any time, hence the solution is global, i.e. $\delta=+\infty$. \ For $\epsilon <0$, in the sub-critical case $n \sigma <2$ then the conservation of the energy still give the same estimate on $\| \nabla \psi \|$ and the global existence of the solution. \ In fact, for $\epsilon <0 $ small enough, by energy conservation and bootstrap arguments (see \cite {Sa2}) any solution is global in the critical $n\sigma =2$ and hyper-critical $n\sigma >2$ cases too.

\subsection{Linear beats}

The operator $H_0$ formally defined by \eqref {eq2} admits a self-adjoint realization (still denoted by $H_0$) on $L^2 ({\mathbb R}^d )$ (Theorem III.1.1 in \cite {BS}) with purely discrete spectrum $\sigma (H_0 ) = \sigma_d (H_0)$, where $\sigma_d (H_0)$ denotes the discrete spectrum. \ Let $\lambda_k$, $k \in {\mathbb N}$, be the eigenvalues of $H_0$
\begin{eqnarray*}
\lambda_1 < \lambda_2 < \lambda_3 \le \lambda_4 \le \ldots \le \lambda_k \le \ldots 
\end{eqnarray*}
with associated normalized (in $L^2$) eigenvectors $\varphi_k (x)$. \ In particular, the lowest two eigenvalues of $H_0$ are non-degenerate and there exists $0<C<1$, independent of $\hbar$, such that
\begin{eqnarray*} 
1+ C \hbar < \lambda_{1,2} < 1+C^{-1} \hbar \ \ \mbox { and } \ \ \inf_{\lambda \in \sigma (H_0) - \{ \lambda_1, \ \lambda_2 \} } [\lambda - \lambda_{1,2} ] \ge C \hbar 
\end{eqnarray*}
and
\begin{eqnarray}
\lambda - \lambda_{2} \ge C \lambda \hbar , \ \forall \lambda \in \sigma (H_0) - \{ \lambda_1, \ \lambda_2 \} \ .
\label {eq6}
\end{eqnarray}
It is well known that the \emph {splitting} between the two lowest eigenvalues
\begin{eqnarray}
\omega = \frac 12 (\lambda_2-\lambda_1) \label {eq7}
\end{eqnarray}
vanishes as $\hbar $ goes to zero. \ In order to give a precise estimate of the splitting $\omega$ let
\begin{eqnarray*}
\Gamma = \inf_\gamma \int_{\gamma} \sqrt {V (x)-V_{min}} d x >0
\end{eqnarray*}
be the Agmon distance between the two wells; where $\gamma$ is any path connecting the two wells, that is $\gamma \in AC ([0,1],{\mathbb R}^d)$ such that $\gamma (0)=x_-$ and $\gamma (1)=x_+$. \ From standard WKB arguments (see \cite {He}) it follows that the splitting is \emph
{exponentially small}, precisely that
\begin{eqnarray}
\omega = \tilde \asy ( e^{- \Gamma /\hbar } ). \label {eq8}
\end{eqnarray}

The normalized eigenvectors $\varphi_{1,2}$ associated to $\lambda_{1,2}$ can be chosen to be real-valued functions such that $\varphi_1$ and $\varphi_2$ are respectively of even and
odd-parity:
\begin{eqnarray*}
\varphi_{j} (-x_1, x_2 , \ldots , x_d ) = (-1)^{j+1} \varphi_{j} (x_1, x_2 , \ldots , x_d ), \ \ j=1,2; 
\end{eqnarray*}

We define now the \emph {single well states}
\begin{eqnarray*}
\varphi_R = \frac {1}{\sqrt 2} \left [ \varphi_1 + \varphi_2 \right ] \ \ \mbox { and } \ \ \varphi_L = \frac {1}{\sqrt 2} \left [ \varphi_1 - \varphi_2 \right ]
\end{eqnarray*}
such that 
\begin{eqnarray*}
\varphi_{R } (-x_1,x_2,\ldots ,x_d)=\varphi_{L}(x_1,x_2 , \ldots , x_d) 
\end{eqnarray*}
They are \emph {localized on one well} in the sense that for any $r>0$ there exists $C_r>0$ such that 
\begin{eqnarray*}
\int_{D_r(x_+)} |\varphi_{R} (x)|^2 dx = 1 + \asy (e^{-C_r /\hbar }) 
\end{eqnarray*}
and
\begin{eqnarray*} 
\int_{D_r(x_-)} |\varphi_{L} (x)|^2 dx = 1 + \asy (e^{-C_r /\hbar })  
\end{eqnarray*}
where $D_r(x_\pm )$ is the ball with center $x_\pm$ and radius $r$. \ For such a reason we call them \emph {single-well} (normalized) states. \ In particular
\begin{eqnarray*}
\| \varphi_{R } \varphi_{L} \|_{L^\infty} = \tilde \asy (  e^{-\Gamma /\hbar })
\end{eqnarray*}

Let 
\begin{eqnarray*} 
\Pi = \langle \varphi_1 , \cdot \rangle_{L^2} \varphi_1 + \langle \varphi_2 , \cdot \rangle_{L^2} \varphi_2  \ \ \mbox { and } \ \ \Pi_c = {\mathbb I} - \Pi
\end{eqnarray*}
be the projection operator onto the eigenspace orthogonal to the bi-dimensional space associated to the doublet $\{ \lambda_{1,2} \}$. We will study the dynamics of equation \eqref{eq1} with initial data almost exactly concentrated on $\Phi_0:=$Span$(\varphi_1,\varphi_2)$. 

\begin{remark} \label{Rmk3}
If one takes initial data $\psi^0$ in $\Phi_0$, i.e. of the form 
\begin{eqnarray*} 
\psi^0 = \zeta_1 \varphi_1 + \zeta_2 \varphi_2 = \zeta_R \varphi_R + \zeta_L \varphi_L , \ \
\zeta_R = \frac {\zeta_1+\zeta_2}{\sqrt 2}, \ \zeta_L = \frac {\zeta_1-\zeta_2}{\sqrt 2} 
\end{eqnarray*}
then the \emph {linear} Schr\"odinger equation
\begin{eqnarray*}
\left \{
\begin {array}{l}
i \hbar \dot \psi^t = H_0 \psi^t , \\
\left. \psi^t  (x) \right |_{t=0} = \psi^0 (x) \in L^2 ({\mathbb R}^d ) , \ \Pi_c \psi^0  =0 , 
\end {array}
\right. 
\end{eqnarray*}
has an explicit solution given by
\begin{eqnarray*}
\psi^t (x) = e^{-i \Omega t /\hbar} \left [  \left ( \zeta_R \varphi_R + \zeta_L \varphi_L \right ) \cos (\omega t/\hbar ) +  i \left ( \zeta_L \varphi_R + \zeta_R \varphi_L \right )\sin (\omega t / \hbar ) \right ] 
\end{eqnarray*}
where
\begin{eqnarray}
\Omega=\frac{\lambda_1+\lambda_2}{2}\ ,\quad \omega=\frac{\lambda_2-\lambda_1}{2} \label {eq9}
\end{eqnarray}
That is $\psi^t (x)$ performs a \emph {beating motion} with \emph {beating period}
\begin{eqnarray*}
T= \frac {2\pi \hbar}{\omega}
\end{eqnarray*}
Such a period usually plays the role of unit of time. 
\end{remark}

\subsection{The nonlinear system}

To introduce the analytic framework in which we will work we first give the following
 
\begin{definition} 
For any integer number $s \ge 0$ we define the Hilbert space ${\mathcal X}_s:=D(H_0^{s/2})$ equipped with the norm
\begin{eqnarray*}
\norma{\psi}_s^2:=\norma{H_0^{s/2}\psi}_{L^2}^2\equiv \left\langle H_0^s\psi,\psi \right\rangle_{L^2}
\end{eqnarray*}
\end{definition}

\begin{remark} \label{Rmk4}
Here we will use the energy space ${\mathcal X}_1$. \ In particular, the  Gagliardo-Niremberg inequality yields the following 
\begin{eqnarray*}
\norma{\psi}_{L^{2\sigma+2}} \leq \frac {C}{\hbar^{e_\sigma}} \| (-\hbar^2 \Delta )^{1/2} \psi \|_{L^2}^{e_\sigma} \cdot 
\| \psi \|_{L^2}^{1-e_\sigma}
\end{eqnarray*}
where $\| \psi \|_{L^2} =1$ and $e_\sigma$ a suitable positive number. \ Hence, we can conclude that 
\begin{eqnarray*}
\| \psi \|_{L^{2\sigma+2}} \le \frac {C}{\hbar^{e_\sigma}} \| \psi \|_1^{e_\sigma}
\end{eqnarray*}
since
\begin{eqnarray*}
\left \| \left (-\hbar^2 \Delta \right )^\frac 12 \psi \right \|_{L^2}^2 = \langle \psi , - \hbar^2 \Delta \psi \rangle \le \langle \psi , H_0 \psi \rangle = \| \psi \|_1^2 .
\end{eqnarray*}
\end{remark}

\section{Main Results} \label{s.3}

\begin {hypothesis} 
Let $\omega$ be the splitting \eqref{eq7} (which satisfies the asymptotic estimate \eqref {eq8}), and let $\epsilon$ be the strength of the non-linear term. \ We assume that the real-valued parameter $\epsilon$ depends on $\hbar$ in such a way
\begin{eqnarray} 
\frac {|\epsilon |\hbar^{-d\sigma /2}}{\omega} \le C, \ \ \forall \hbar \in (0 , \hbar^\star ) \label {eq10} 
\end{eqnarray}
for some positive constant $C$, independent of $\hbar$, and for some $\hbar^\star$.
\end {hypothesis}

\begin {remark} The ratio
\begin{eqnarray*}
\eta = \frac {\epsilon \hbar^{-d\sigma /2}}{\omega}
\end{eqnarray*}
plays the role of effective nonlinearity parameter. \ The above assumption implies that $|\eta | \le C$.
\end {remark}

The main result of this section is the following 

\begin{theorem} \label{Thm1}
Assume Hypotheses 1,2, and consider the Cauchy problem \eqref{eq1}--\eqref{eq3}. \ Then there exist positive constants $0<C<1$ and $\gamma$ such that, if $\hbar$ is small enough and the initial datum $\psi^0$ fulfills
\begin{eqnarray}
\label{eq11} \norma{\Pi_c\psi^0}_1 \le C {|\epsilon|^{1/2}} \ ,
\end{eqnarray} 
then one has
\begin{eqnarray}
\label{eq12}
\norma{\Pi_c\psi^t }_1 \le \frac{|\epsilon|^{1/2}}{C\hbar^{\gamma}} ,\quad \forall t\in{\mathbb R} \ .
\end{eqnarray}
\end{theorem}

Hence the 2-dimensional space $\Phi_0:=\Pi \left [ L^2 ({\mathbb R}^d ) \right ] $ is almost invariant. \ Thus one expects that corresponding to initial data satisfying \eqref{eq11} the dynamics is well described by the restriction of the equations of motion to $\Phi_0$. \ Actually such a restricted dynamical system coincides, up to (formal) order $\epsilon^2$, with the 2-dimensional
dynamical systems
\begin{eqnarray} \label {eq13}
\left \{
\begin {array}{ll}
i \hbar \dot c_R = - \omega c_L + \Omega c_R + \epsilon C_\sigma |c_R|^{2\sigma } c_R   \\ 
i \hbar \dot c_L = \Omega c_L - \omega c_R + \epsilon C_\sigma |c_L|^{2\sigma } c_L  
\end {array}
\right. , 
\end{eqnarray}
where 
\begin{eqnarray*}
 C_\sigma = \| \varphi_R^{2\sigma} \|^2_{L^2} = \| \varphi_L^{2\sigma} \|_{L^2}^2 = \asy (\hbar^{-d\sigma /2} ),
\end{eqnarray*}
which has the integral of motion 
\begin{eqnarray*}
{\mathcal I} (c_R , c_L ) =\Omega (|c_R|^2 + |c_L|^2) - \omega \left ( \bar c_R c_L + \bar c_L c_R \right ) + C_\sigma \frac {\epsilon}{\sigma +1} \left ( |c_R|^{2(\sigma +1)}+ |c_L|^{2(\sigma +1)} \right ) \ .
\end{eqnarray*}
Moreover, such an Hamiltonian system has an independent integral of motion (the quadratic part of the Hamiltonian), and thus it is integrable. \ The 2--dimensional system was studied in detail in \cite {GMS} obtaining that when the nonlinearity parameter $\eta = \frac {\hbar^{-d\sigma /2}\epsilon}{\omega}$ is large enough almost all its solution do not posses the beating property, i.e. the probability density remains concentrated in one well. 

Concerning the relation between the solution of the two dimensional system and the solution of the complete system we have the following 

\begin{corollary} \label{Cor1}
Let $\varphi^a(t) = C_R \varphi_R + C_L \varphi_L$, where $C_R$ and $C_L$ are the solution of the 2--dimensional system \eqref {eq13} with initial datum $\varphi^a (0)$ with of norm 1, i.e. $|c_R|^2+|c_L|^2 =1$. \ Denote $\varphi(t):=\Pi \psi(t)$, where $\psi(t)$ is the solution of the complete system \eqref{eq1} with the same initial datum. \ Then there exist a positive $C$ and a positive $\alpha$ such that the following estimate holds:
\begin{eqnarray*}
\norma{\varphi^a(t)-\varphi(t)}_{L^2}\leq C\frac{\epsilon^{3/2}}{\hbar^\alpha}|t| \ .
\end{eqnarray*}
\end{corollary}

\section {Proof of the main results}

Let
\begin{eqnarray*} 
\psi^t (x) = \sum_{k=1}^\infty \zeta_k (t) \varphi_k (x), 
\end{eqnarray*}
and define the Hilbert spaces $\ell^2_s$ of the complex sequences $\zeta = \left \{ \zeta_m \right \}_{m \in {\mathbb N}}$ such that
\begin{eqnarray*}
\norma{\zeta}_s^2:=\sum_{k\geq1}\lambda_k^s|\zeta_k|^2<\infty
\end{eqnarray*}
and remark that in such a way we have defined the correspondence 
\begin{eqnarray*}
\psi \in {\mathcal X}_s \leftrightarrow \zeta = {\mathcal U} (\psi ) \in \ell^2_s
\end{eqnarray*}
which is a unitary isomorphism.

In terms of these variables the norm and quadratic part ${\mathcal E}_0$ of the Hamiltonian are given by
\begin{eqnarray*}
{\mathcal N} = \sum_{k \ge 1} |\zeta_k|^2 = \| \zeta \|^2_0\ \ \mbox { and } \ \ {\mathcal E}_0 =\sum_{k\geq1} \lambda_k \left | \zeta_k \right|^2\ =\| \zeta \|_1^2 .
\end{eqnarray*} 

\begin {remark} 
In order to simplify the notations, {\bf we rescale time by the transformation} $t\mapsto t/\hbar$. \ With such a notation the beating period is now given by 
\begin{eqnarray*}
T= \frac {2\pi}{\omega}
\end{eqnarray*}
\end {remark}

\subsection{Variational results}

We recall that we work here in the energy space ${\mathcal X}_1$. \ Thus, in this section, norms and distances will always be in this space.

To exploit the fact that both ${\mathcal E}_\epsilon$ and ${\mathcal N}$ are conserved along the flow we have to introduce a few geometrical lemma.

Denote 
\begin{eqnarray*}
{\mathcal H}_0(\psi):=\Omega(|\zeta_1|^2+|\zeta_2|^2)+\sum_{k\geq 3} \lambda_k \left | \zeta_k \right| ^2 = \sum_{k \ge 1} \nu_k |\zeta_k |^2
\end{eqnarray*}
and
\begin{eqnarray*}
{\mathcal P}_{\epsilon}(\psi ):= \omega(|\zeta_2|^2-|\zeta_1|^2)+\epsilon P(\psi )\ ,
\end{eqnarray*}
where $\nu_1 = \nu_2 = \Omega$, $\nu_k = \lambda_k$ for $k \ge 3$ and $\Omega$ and $\omega$ are defined in \eqref {eq9}, so that ${\mathcal E}_\epsilon={\mathcal H}_0+{\mathcal P}_\epsilon$. 

In this section we will almost always use real coordinates $p_j,q_j$ defined by
\begin{eqnarray*}
\zeta_j=\frac{q_j+i p_j}{\sqrt2}\ ,
\end{eqnarray*}
so all the functions will be considered as functions of $\psi$, $(p,q)$, or $\zeta_j$ according to convenience. 

Define the (smooth) surface 
\begin{eqnarray*}
{\mathcal S}:=\left\{\psi\in{\mathcal X}_1\ :\ {\mathcal N}(\psi )=1\right\}\ ,
\end{eqnarray*}
and consider the function $h_0:={\mathcal H}_0\big|_{{\mathcal S}}$.

\begin{lemma}
The manifold
\begin{eqnarray*}
&& N:={\mathcal S} \cap \Pi \left ( L^2 ({\mathbb R}^d )\right )  = \\
&& \nonumber \ = \left\{\psi\in{\mathcal X}_1\ :\ \psi=\zeta_1\varphi_1+\zeta_2\varphi_2\
,\quad |\zeta_1|^2+|\zeta_2|^2=\frac{p_1^2+q_1^2+p_2^2+q_2^2}{2}=1\right\}
\end{eqnarray*}
is an absolute minimum of $h_0$. \ Furthermore, for any point $\psi \in N$ decompose the tangent space $T_\psi {\mathcal S}$ as
\begin{eqnarray*}
T_\psi{\mathcal S}= T_\psi N\oplus (T_\psi N)^{\perp }.
\end{eqnarray*}
Then, the second differential ${\rm d}^2_\psi h_0$ of $h_0$ at a point $\psi\in N$ is such that 
\begin{eqnarray}
\label{eq14}
{\rm d}^2_\psi h_0(X,X)\geq C \hbar \norma{X}_1^2\ ,\quad \forall X\in (T_\psi N)^{\perp } 
\end{eqnarray}  
\end{lemma}

\begin{remark} \label{Rmk7}
Since the function $h_0$ is smooth one also has
\begin{equation}
\label{eq15}
{\rm d}^2_\psi h_0(X,X)\leq C\norma{X}_1^2\ ,\quad \forall X\in (T_\psi N)^{\perp } 
\end{equation}
with a suitable $C$.
\end{remark}

\proof It is a trivial application of the method of the Lagrange multipliers. \ Consider ${\mathcal H}_0+\lambda {\mathcal N}$; the critical points of $h_0$ are obtained by finding the zeros (on ${\mathcal S}$) of the equations 
\begin{eqnarray}
\label{eq16}
\frac{\partial({\mathcal H}_0+\lambda{\mathcal N})}{\partial p_j}= \frac{1}{2}(\nu_j+\lambda)p_j  \ ,
\end{eqnarray}
where $\nu_j=\Omega$ for $j=1,2$ and $\nu_j=\lambda_j$ for $j\geq 3$, and of the analogous equation for $q_j$. \ Thus a solution of \eqref{eq16} is given by $\lambda=-\Omega$ and $\psi\in N$. \ It follows that $N$ is constituted by critical points of $h_0$. \ It follows that the second differential of $h_0$ at such points is well defined. \ Moreover, we recall that, given a vector 
\begin{eqnarray*}
Y=\sum_{j\geq 1}(Q_j+i P_j)\varphi_j\in T_\psi{\mathcal S} , 
\end{eqnarray*}
one has 
\begin{eqnarray*}
&&{\rm d}^2_\psi h_0(Y,Y) \\
&&=\sum_{j,l}\left(P_jP_l\frac{\partial^2({\mathcal H}_0+\lambda{\mathcal N}) }{\partial P_j\partial P_l}+
2P_jQ_l\frac{\partial^2({\mathcal H}_0+\lambda{\mathcal N}) }{\partial P_j\partial Q_l}+Q_jQ_l\frac{\partial^2({\mathcal H}_0+\lambda{\mathcal N}) }{\partial Q_j\partial Q_l} \right) \\
&&= \sum_{j\geq 1}(\nu_j+\lambda)\left(P_j^2+Q_j^2 \right) 
\end{eqnarray*}
where $\lambda =-\Omega $ is the Lagrange multiplier determined from the criticality condition. \ By the condition $Y\in (T_\psi N)^{\perp }$ one has $P_j=Q_j=0$ for $j=1,2$ and thus one has 
\begin{eqnarray*}
{\rm d}^2_\psi h_0(Y,Y)\geq \sum_{j\geq3} C \hbar  \left( P_j^2+Q_j^2  \right)= C \hbar \norma{Y}_1^2\ ,\quad \forall
Y\in(T_\psi N)^{\perp} 
\end{eqnarray*}
since \eqref {eq6}, and therefore the thesis on the second differential follows. \ In particular $N$ is a minimum of $h_0$. In a similar way one can show
that all the other critical points of $h_0$ are saddle points. \qed

\begin {remark}
By definition it follows that
\begin{eqnarray*}
\left. h_0 \right |_N \equiv \Omega .
\end{eqnarray*}
\end {remark}

\begin{lemma} \label{Lem2}
There exists a positive $C$ such that, provided 
\begin{eqnarray}
h_0(\psi)-\Omega <\frac {1}{C}\hbar^\beta, \ \ \mbox { for some } \ \beta >3, \label {eq17}
\end{eqnarray}
then one has
\begin{eqnarray}
\label{eq18}
\frac{\hbar }{C}\left[d(\psi,N)\right]^2\leq h_0(\psi)-\Omega \leq C  \left[d(\psi,N)\right]^2 \ . 
\end{eqnarray}
\end{lemma}
\proof This is a standard result in differential geometry; we give here its simple proof for the sake of completeness. \ Actually it is based on the use of the exponential coordinates (see appendix \ref{appA} for their construction), which are coordinates 
\begin{eqnarray*}
\psi \to (n,w)\in N\times (T_{(n,0)}N)^\perp
\end{eqnarray*} 
in which the intersection of $N$ with the domain of definition of the coordinates coincides with $w=0$; moreover for any $\psi=(n,w)$ one has
\begin{eqnarray*}
d (\psi , N) := d\left((n,w);N\right)=\norma{w}_1\ .
\end{eqnarray*}
where the distance is as usual defined as the length of the shortest geodesic from $\psi$ to $N$. \ Using these coordinates, consider the Taylor expansion in $w$, at a point of $N$, of the function
\begin{eqnarray}
\label{eq19}
h_0(n,w)-h_0(n,0)={\rm d}^2_{(n,0)}h_0(w,w)+\asy (\norma{w}_1^3)\ , 
\end{eqnarray}
where $h_0 (n,w ) = h_0 (\psi )$ and $h_0 (n,0) \equiv \Omega \equiv \left. h_0 \right |_N$. \ From this fact and from \eqref{eq14} and \eqref {eq17} then  the left hand side inequality \eqref {eq18} follows. \ The right hand side inequality of \eqref {eq18} follow from \eqref {eq15} and \eqref {eq19}. \qed 

\begin {corollary} \label {Corr2} Let $\psi \in {\mathcal S}$ such that $\| \Pi_c \psi \|_1 = \mu = \asy (\hbar^{\beta /2}) $ for some $\beta >3$, then 
\begin{eqnarray}
\label {eq20} 
C \hbar^{\frac 12} \mu \le d (\psi , N) \le C \hbar^{-\frac 12} \mu 
\end{eqnarray}
for some $C>0$.
\end {corollary}

\begin {proof}
Indeed, since
\begin{eqnarray*}
h_0 (\psi ) &=& \Omega \left ( |\zeta_1 |^2 + |\zeta_2 |^2 \right ) + \sum_{k \ge 3} \lambda_k |\zeta_k |^2 \\ 
&=& \Omega \left ( |\zeta_1 |^2 + |\zeta_2 |^2 \right ) + \| \Pi_c \psi \|_1^2  \\ 
&=& \Omega - \Omega \left \| \Pi_c \psi \right \|_0^2 + \left \| \Pi_c \psi \right \|_1^2 
\end{eqnarray*}
then 
\begin{eqnarray*}
h_0 (\psi ) - \Omega = \| \Pi_c \psi \|_1^2 - \Omega \| \Pi_c \psi \|_0^2 \le \| \Pi_c \psi \|_1^2 =  \mu^2 \ .
\end{eqnarray*}
From this fact and from Lemma \ref {Lem2} the right hand side inequality \eqref {eq20} follows. \ Similarly
\begin{eqnarray*}
h_0 (\psi ) - \Omega = \sum_{j\ge 3} (\lambda_j - \Omega ) | \zeta_j |^2 \ge C \hbar \sum_{j\ge 3} \lambda_j | \zeta_j |^2 \ge C \hbar \| \Pi_c \psi \|_1^2  
\end{eqnarray*}
from which the left hand side inequality \eqref {eq20} follows.
\end {proof}

{\bf Proof of theorem \ref{Thm1}.} First remark that $\psi^0\in{\mathcal S}$ and $\norma{\Pi_c\psi^0}_1=\mu = \asy (\epsilon^{1/2} )$ implies (see Corollary \ref {Corr2}) $d(\psi^0,N)<C \hbar^{-1/2}\mu = \tilde \asy (\epsilon^{1/2} )$. \ Then, remark that from \eqref {eq10} then 
\begin{eqnarray*}
{\mathcal E} (\psi^0) &=& \| \psi_0 \|_1^2 + \frac {\epsilon}{1+\sigma } \| \psi_0^{\sigma +1} \|^2_{L^2} \\ 
&=& \| \Pi \psi_0 \|_1^2 + \asy ( \epsilon^{1/2} ) = \Omega + \asy ( \epsilon^{1/2} )
\end{eqnarray*}
and that, by conservation of energy and Remark \ref {Rmk4} one has $\norma{\psi(t)}_1\leq 2\Omega $ for all times. \ From this fact and from Remark \ref {Rmk4} one also has the a priori estimate which holds for dimension $d=1$ and $d=2$
\begin{eqnarray*}
\left|{\mathcal P}_\epsilon(\psi(t))\right|\leq C\frac{\epsilon}{\hbar^{\beta} } 
\end{eqnarray*}
for some $\beta$; thus, using \eqref{eq18} one has the chain of inequalities 
\begin{eqnarray*}
&& \left[d^2(\psi(t),N)\right]^2 \leq  \frac {C}{\hbar}[h_0(\psi(t))-\Omega ] \\
\nonumber && \ \ \leq \frac {C}{\hbar} \left [ \left ( h_0(\psi^0)-\Omega \right )+ \left ( {\mathcal E}_\epsilon (\psi (t)) - {\mathcal E}_\epsilon (\psi^0) \right ) +\left ( {\mathcal P}_\epsilon
(\psi^0)-{\mathcal P}_\epsilon (\psi(t))\right ) \right ] \\ 
\nonumber && \ \ \leq \frac {C}{\hbar} \left [ \left ( h_0(\psi^0)-\Omega \right ) +\left ( {\mathcal P}_\epsilon
(\psi^0)-{\mathcal P}_\epsilon (\psi(t))\right ) \right ] \le \frac{C}{\hbar^{\beta +1}} \epsilon \ .
\end{eqnarray*}
From this fact and from \eqref {eq20} then the thesis follows. \qed

Finally, Corollary \ref {Cor1} immediately follows by comparing the two-level approximation \eqref {eq13} with \eqref {eq1} and by means of estimate \eqref {eq12} and standard Gronwall's Lemma arguments (see e.g. \cite {Sa2}). 

\appendix

\section{Construction of exponential coordinates}
\label{appA}

Let $M$ and $N$ be Riemannian manifolds modeled on Hilbert spaces $H$
and $K$, with norms $\| \cdot \|_H$ and $\| \cdot \|_K$; denote by $g$ the metric of $M$; let $i:N\to M$ be a smooth
isometric embedding. \ In the following, for the sake of simplicity, we
will identify $N$ and with $i(N)$, and similarly for related objects
and spaces (as $TN$).

\begin{lemma}
Let $x_0$ be a point of $N$ and let $W=(T_{(x_0)}N)^{\perp}$, then there exists a coordinate system of
$M$ defined in a neighborhood ${\mathcal U} \subset M$ of $x_0$ with the following properties:
\begin{itemize}

\item[i.] ${\mathcal U}\ni x\mapsto (n,w)\in  N\times W $;

\item[ii.] let $ d(x,N)$ be the distance usually defined as the length of the shortest geodesic from $x \in N$ to $N$; then $d((n,w);N)=\norma{w}_K$. 

\end{itemize}
\end{lemma}
\proof We proceed in some steps. \ To start with we choose a coordinate system $(n,w_1)$ with origin $x_0 \in N$ such that, if $(n,w_1)\in {\mathcal V}_1\subset N\times W$, where ${\mathcal V}_1$ is a neighborhood of $x_0$; then the intersection of $N$ with the domain of definition of the coordinate system coincides with the set $(n,0)$. \ Let 
\begin{eqnarray*}
\Pi_n: {\mathcal V}_1 \subset N\times W\to \left(T_{(n,0)}N\right)^\perp
\end{eqnarray*}
be the orthogonal projector with respect to the scalar product $g_{(n,0)}$. \ Define the map 
\begin{eqnarray*}
{\mathcal V}_1\ni (n,w_1)\mapsto (n, \Pi_n w_1 )\in{\mathcal V} \ .
\end{eqnarray*} 
If ${\mathcal V}_1$ is small enough, then such a map is an isomorphism on its image, since its differential at the origin is invertible. \ Thus it defines a new coordinate system. \ Let $(n,0) \in V_1$, then in these coordinates one has
\begin{eqnarray*}
T_{(n,0)}N=\left\{(X_1,X_2)\in N\times W\ :\ X_2=0\right\} \\
(T_{(n,0)}N)^\perp=\left\{(X_1,X_2)\in N\times W\ :\ X_1=0\right\}
\end{eqnarray*}
We use such a coordinate system in order to define the needed coordinate system. Take $(n,w)\in N\times W$ small enough, and
consider the geodesic $\gamma_{(n,w)}(s)$ starting from $(n,0)$ with initial velocity $w$. \ Consider the (exponential) map
\begin{eqnarray*}
(n,w)\mapsto \gamma_{(n,w)}(1)\ ,
\end{eqnarray*} 
by implicit function theorem it is locally invertible (its differential at the origin is the identity), and thus any point $x$ of a neighborhood ${\mathcal U}$ of $x_0$ can be represent uniquely by the points $(n,w)$ such that $\gamma_{(n,w)}(1)=x_0$. \ Remark that moreover one has
\begin{eqnarray}
\label{eq21}
\ell(\gamma):=\int_0^1\sqrt{g(\dot\gamma(s),\dot\gamma(s))}ds=\norma{w}_K 
\end{eqnarray}
since $\dot \gamma(0)=\norma{w}_K$ and $g(\dot\gamma(s),\dot\gamma(s))$ is independent of $s$ along geodesic lines. 

In the coordinate system just introduced, let $x=(n,w)$ then its distance from $N$ is the length of the shortest geodesic joining $x$ to
$N$. \ In turn such a geodesic is perpendicular to $N$. Thus $(n,0)$ is the point where it starts and $w$ is the tangent vector at such a
point. \ Then property ii. follows from \eqref{eq21}.\qed

\end {document}